\documentclass[12pt]{iopart}
\usepackage{graphicx}

\begin{document}
\title[Current saturation and Coulomb interactions in organic single-crystal 
transistors]
{Current saturation and Coulomb interactions in organic single-crystal 
transistors}

\author{S. Fratini$^1$, H. Xie$^2$, I. N. Hulea$^2$,
S. Ciuchi$^3$, and A. F. Morpurgo$^2$}

\address{$^1$ Institut N\'eel - CNRS \& Universit\'e Joseph Fourier,
BP 166, F-38042 Grenoble Cedex 9, France}
\address{$^2$  Kavli Institute of Nanoscience, Delft University of Technology,
Lorentzweg 1, 2628 CJ Delft, The Netherlands} 
\address{$^3$ INFM and Dipartimento di Fisica
Universit\`a dell'Aquila,
via Vetoio, I-67010 Coppito-L'Aquila, Italy}

\begin{abstract}
Electronic transport through rubrene single-crystal field effect
transistors (FETs) is investigated experimentally in the high carrier density regime ($n \simeq 0.1$ carrier/molecule). In this regime, we find that the current does not increase linearly with the density of charge carriers, and
tends to saturate. At the same time, the activation energy for
transport unexpectedly increases with increasing $n$. We perform a
theoretical analysis in terms of a well-defined microscopic model for
interacting Fr\"ohlich polarons, that quantitatively 
accounts for our experimental observations. 
This work is particularly significant for our understanding of
electronic transport through organic FETs. 
\end{abstract} 
\submitto{\NJP}
\maketitle

\section{Introduction}
The use of single-crystalline material for the fabrication of organic 
field-effect transistors (FETs) has given, over the past few years,
the experimental control needed for the investigation of the 
intrinsic transport properties of dielectric/organic interfaces
\cite{RevModPhys}.   
This has resulted in the observation of anisotropic transport  \cite{Sundar}, 
a metallic-like temperature dependence of the mobility \cite{Podzorov1},
Hall-effect \cite{Podzorov2}, and quasiparticle response in the
infrared conductivity \cite{Basov}. 
Following this progress, the successful quantitative analysis of experiments
in terms of a simple microscopic model
has  been recently possible in single-crystal FETs with highly 
polarizable gate dielectrics \cite{NatMat}. 
In these devices, charge carriers accumulated electrostatically at the
surface of the organic semiconductor were shown to interact strongly
with the polarization in the dielectric, leading to the formation of
Fr\"ohlich polarons \cite{Kirova}.  The
quantitative agreement between the behavior predicted by 
the Fr\"ohlich  Hamiltonian and the experimental data demonstrates that 
the microscopic processes responsible for transport in  organic FETs
can be qualitatively  different than in bulk organic semiconductors.  

Here we investigate the transport properties of rubrene
(C$_{42}$H$_{28}$) single-crystal FETs with highly polarizable
Ta$_2$O$_5$ gate dielectrics, in the yet unexplored high carrier
density regime.   We find that in this regime the electrical characteristics of
the devices exhibit pronounced deviations from 
those of conventional FETs. Specifically, in all devices the
source-drain current $I_{sd}$ stops increasing linearly with the gate
voltage $V_g$, and shows a clear saturation. 
Concomitantly,
the activation energy for temperature-dependent transport increases
when $V_g$ increases, a trend opposite to that usually
observed in organic transistors. We build on the Fr\"ohlich model, that
was used in our previous study  to describe the dressing of the
carriers by the polarizability of the gate dielectric \cite{NatMat},
and find that the observed behavior can be quantitatively explained by
considering the effects of the Coulomb interactions between
holes. Our results confirm that organic single-crystal transistors are
suitable for the experimental  investigation of the intrinsic
electronic properties of dielectric/organic interfaces, and 
extend our fundamental understanding of transport in organic
transistors to the high carrier density regime.

\section{Experimental results}
The rubrene transistors used in our experiments are fabricated by
laminating thin ($< 1 \mu$m thick) rubrene single-crystals onto a
substrate with pre-fabricated FET circuitry, as described in
Refs. \cite{deBoer, Review}. The gate dielectric is a $\simeq 500$ nm-thick layer of Ta$_2$O$_5$ (dielectric constant $\epsilon_s=25$; breakdown voltage
$\simeq 6.5 MV/cm$), enabling the accumulation of up to $10^{14}$ holes/cm$^2$ in the FET channel. During device operation, however,
we never exceed gate voltage values corresponding to half the
breakdown field, to minimize the chance of device failure \cite{leakage} (we reach hole densities of 5$\cdot$ 10$^{13}$ cm$^{-2}$, i.e. one order of magnitude higher than in Ref. \cite{NatMat}). The
transistor electrical characteristics were measured in the vacuum
($10^{-6}$ mbar) chamber of a flow cryostat, between 300 K and 125-210
K depending on the specific device (at low temperature the devices can
easily break due to the difference in thermal expansion between
crystal and substrate).

\begin{figure}
  \centering
   \includegraphics[width=90mm]{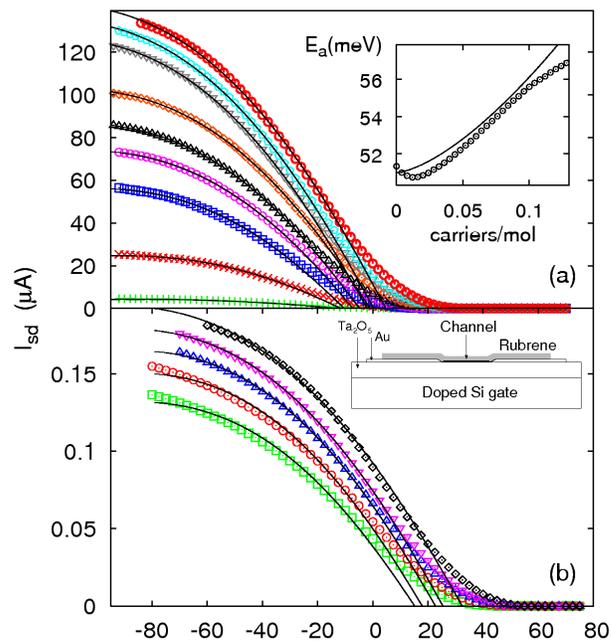}
  \caption{Source-drain current {\it vs.} gate voltage measured for two different
    devices at different temperatures: a) from top to bottom,
    $T=280, 265, 250, 235, 215, 200, 175, 150, 125$ K (measurements
    were performed at a source-drain bias $V_{sd}=-16$ V; the channel
    width and length are $W=340 \mu$m and $L=200 \mu$m, respectively);
    b) from top to bottom, $T=300,260,240, 220, 210$ K ($V_{sd}=-1$ V,
    $W= 28 \mu$m, $L=630 \mu$m). 
    The full lines are the theoretical
    fits.  The inset in a) shows the experimentally measured activation
    energy of $I_{sd}$ plotted {\it
      vs.} carrier density $n=C|V_g-V_{th}|/e$, compared with the
    theory (full line). The experimental error is comparable with the size of
    the data points.
The device is sketched in panel b). }
  \label{fig:Isd}
\end{figure}

Figure 1 shows the source-drain current $I_{sd}$ measured on
two different devices as a function of gate voltage $V_g$, at
different temperatures. Since the carrier density $n$ in the channel
is linearly proportional to the gate voltage $V_g$
[$n=C(V_g-V_{th})/e$, with $C$ capacitance per unit area and $V_{th}$
threshold voltage], it is normally expected that $I_{sd}$ also
increases linearly with $V_g$. The data, however, show a pronounced
deviation from the conventional linear behavior, since at high $V_g$
the source-drain current tends to saturate. The effect is reversible
and reproducible: the devices can be measured many times without
noticeable difference in the data, which implies that the saturation
of $I_{sd}$ is not due to device degradation \cite{leakage}. We
have measured tens of similar 
devices (room-temperature mobility values between 1.0 and 1.5
cm$^2$/Vs) and found a similarly pronounced saturation of the 
source-drain current at high gate voltage in virtually all cases, albeit with
some differences in the details of the $I_{sd}-V_g$ characteristics
(illustrated by the data in Fig.1a and b). 

The temperature dependence of the current $I_{sd}$ is also unusual. At
any fixed value of carrier density $n$, the current $I_{sd}$ decreases
in a thermally activated way \cite{NatMat}. The inset of Fig.1a shows that the
activation energy $E_a$ depends on $n$ and that, for densities larger
than $\simeq 0.02$ holes/molecule, $E_a$ {\it increases with
increasing $n$}. This observation is surprising, because normally
e.g. in organic thin-film FETs) $E_a$ exhibits the opposite trend,
which is attributed to the effect of disorder (filling of traps) in
the organic material.

A behavior of the $I_{sd}$ vs. $V_g$ curves such as the one just described
has not been reported earlier in FETs realized on other conventional gate dielectrics (e.g., SiO$_2$, Al$_2$O$_3$, Si$_3$N$_4$). Only recently, a strongly non-linear $I_{sd}-V_g$ relation was observed in organic FETs with gate
electrolytes \cite{Shimotani, Frisbie}. Gate electrolytes enable the
accumulation of large density of carriers,  $5 \times 10^{13}$ 
carriers/cm$^2$ or more, similar to our Ta$_2$O$_5$ dielectrics. At
these densities, at which the charges accumulate in
the uppermost molecular layer of the crystal \cite{NatMat},  
the average distance between the carriers is only a few molecules
and the resulting  (bare) 
Coulomb interaction is a few hundreds of
milli-electronvolts, much larger than the thermal energy at room
temperature. This, together with the poor screening associated to the thermally activated  motion of charge carriers, suggests that interactions may be
responsible for the observed anomalous FET characteristics.

\section{Microscopic model}
As we proceed to show, the inclusion of Coulomb 
interactions between holes does indeed successfully account for our experimental 
findings. To introduce the theoretical framework needed for the quantitative description of the device characteristics we recall the conclusions of Ref. \cite{NatMat}, namely that holes in rubrene FETs with
highly polarizable gate dielectrics form small 
Fr\"ohlich polarons due to their strong interaction with the
ionic polarization at the interface. 
This conclusion was
established by analyzing transport measurements in the low density regime, 
performed on FETs with a range of different gate dielectrics. Specifically,
 it was shown that, using gate dielectrics of increasing polarizability, 
the ``metallic-like''  transport previously observed in rubrene
\cite{Podzorov1,Troisi}  is progressively suppressed, 
turning into an activated behavior
characteristic of self-localized particles. The experimentally
observed behavior was consistently explained in terms of the Fr\"ohlich
model   --that describes the polar interaction with interface phonon modes--  
by including explicitly the narrow-band nature of the organic crystal
\cite{NatMat}.  In FETs with Ta$_2$O$_5$ devices, the most polarizable 
dielectric used in our previous work, the interaction with the interface modes 
completely dominates the carrier motion, reducing the mobility at 200
K by two orders of magnitude with respect to isolated rubrene.

Here, we extend the analysis of Ref. \cite{NatMat} to the
high-density regime that had not been analyzed previously, by adding a
term which accounts for the mutual Coulomb interactions between 
the carriers \cite{polcryst}. The resulting Hamiltonian:
\begin{equation}
  \label{eq:H}
  H=-t\sum_{<ij>} (c^+_i c_j +c_j^+c_i) + \frac{K}{2} \sum_\ell  X_\ell^2 + \sum_{i, \ell} g_{i\ell}
  n_i X_\ell  
  +\frac{1}{2} \sum_{i\neq j} n_i
  V_{ij} n_j.
\end{equation}
is the simplest model that catches the main microscopic
mechanisms that determine the transport in our devices: 
finite electronic bandwidth,  polar coupling of the carriers 
with the gate dielectric, and mutual interaction between carriers.
The first three terms describe free holes in a tight-binding scheme
($c^+_i,c_i$ are the corresponding creation and annihilation operators
on site $i$), interacting with lattice deformations $X_\ell$ 
through a non-local coupling $g_{i\ell}$. Physically,  $g_{i\ell}$ represents
the polar interaction between the holes in the two-dimensional 
conducting channel and the phonons at the organic/dielectric 
interface \cite{MoriAndo,NatMat}. 
When this interaction is sufficiently strong, as is the case for the
rubrene/Ta$_2$O$_5$ interface,
it leads to the formation of small polarons, i.e.  self-trapped states
whose characteristic radius is comparable with the lattice spacing. 
This is demonstrated theoretically in Appendix A, 
and is confirmed by the experimental data, which exhibit 
a thermally activated  mobility characteristic of small polarons
(see Ref. \cite{NatMat} and Fig. \ref{fig:parameters} below).
As in Ref. \cite{NatMat},  we
explicitly neglect the kinetic energy of the phonons, which is
appropriate in the adiabatic regime, for which the lattice dynamics is
slow compared to the band electrons. 

The last term ---not considered
previously---  represents the mutual interaction between
holes through the unscreened Coulomb repulsion $V_{ij}$, which is
long-ranged (this contrasts with the Hubbard model,
where a purely on-site interaction is considered, leading to the
appearance of interaction effects only at densities close to one
carrier/molecule).  
In the following we evaluate quantitatively the effects of Coulomb 
interactions on the transport characteristics. We shall only present the
main formulas needed for the comparison with the experiments, 
leaving the full details of the calculations to a further publication.
\footnote{It should be noted that 
the electron-lattice interaction in 
Eq.  (\ref{eq:H})  can mediate an effective attraction between holes, leading 
under specific conditions to the stabilization of bipolarons, 
i.e. two holes bound together by a common 
lattice deformation.  This requires in particular that the
phonon-mediated attraction is sufficiently strong to overcome the
instantaneous Coulomb repulsion between holes. 
This condition can be cast in terms
of the effective dielectric constants of the interface at high and low
frequencies, and  corresponds to  
$\eta=(\epsilon_\infty+\kappa)/(\epsilon_s+\kappa)$ 
being less than a given critical value $\eta_c$.
An accurate estimate $\eta_c=0.131$ was given in Ref. \cite{Verbist} 
for a pure 2D Fr\"ohlich interaction in 
the continuum limit.  Such critical value should be further 
reduced in the present case, because the finite distance $z$ of the
conducting channel to the interface 
suppresses the electron-phonon interaction at large momenta, 
$M_q\sim e^{-qz}/\sqrt{q}$ (see Appendix A). 
For the present 
rubrene/Ta$_2$O$_5$ interface we have $\eta=0.26$, which 
lies safely above the critical value $\eta_c$, 
so that  bipolaron formation can in principle be excluded.}

In the temperature range of the experiment the motion of small Fr\"ohlich polarons at the rubrene/Ta$_2$O$_5$ interface takes place
through successive uncorrelated hops from a given molecule
to a nearest neighbor. In this case, the mobility can
be rigorously determined by analyzing the hopping rate of a given
carrier between two neighboring molecules (say $i=1,2$) \cite{LangFirsov}.
Integrating out the electronic degrees of freedom, 
it is found that the hole dynamics is coupled to the long-range
lattice polarization  solely through 
the collective coordinate $Q=\frac{1}{g}\sum_\ell(g_{1\ell}-g_{2\ell})
X_\ell$.  Taking advantage of the adiabatic assumption, 
we end up with the following double-well potential:
\begin{equation}
  \label{eq:adpot}
  E_{ad}(Q)=\frac{K}{4} Q^2-\sqrt{\left(
      \frac{g}{2}Q-\frac{\xi}{2}\right)^2+t^2}
\end{equation}
where we have defined $g^2=\sum_\ell
g_{1\ell}(g_{1\ell}-g_{2\ell})$.
The physical meaning of the above equation is that in the adiabatic regime, 
the hole follows instantaneously the dynamics of 
the  slow coordinate $Q$  in the effective potential $E_{ad}$.
The two minima at $Q\simeq \mp g/2K$ are then associated with the
hole being at site $1$ and $2$ respectively. The corresponding
hopping rate can be evaluated by standard techniques \cite{Chandra}, by
studying the escape rate of the coordinate $Q$ over the barrier.
The variable $\xi$ that appears in Eq. (\ref{eq:adpot}) (defined
below) accounts for the instantaneous
repulsion of all the other carriers in the conducting layer, and vanishes
in the low density limit.
In that case the double-well potential 
Eq. (\ref{eq:adpot}) is symmetric, with a barrier given 
by $\Delta= g^2/4K -t$ 
(the polaronic gap, valid in the strong coupling regime $g^2/4K \gg t$), 
all the carriers diffuse with the same hopping rate
$\rm{w}_0=(\omega_s/2\pi)e^{-\Delta/k_BT}$,
and we recover  the mobility of independent polarons obtained in
Ref. \cite{NatMat}, namely $\mu_P(T)=(ea^2/\hbar T)\rm{w}_0$ (here 
 $\omega_s$ is the frequency of the interface optical phonons and $a$ is the 
intermolecular distance).

In the high density regime accessible with Ta$_2$O$_5$ gate
dielectrics it becomes essential to include the Coulomb interactions
that were not analyzed previously, which requires considering the case
of a finite $\xi$.  
From the model Eq. (\ref{eq:H}) it can be shown that: 
\begin{equation}
  \label{eq:xi}
  \xi = \sum_{j\neq 1,2} (\bar{V}_{2j}-\bar{V}_{1j}) n_j,
\end{equation}
where $\bar{V}_{ij}=2e^2/(\epsilon_s+\kappa)/R_{ij}$ is the Coulomb
potential, appropriately screened by the ionic polarization at the interface 
($\kappa=3$ is the dielectric constant of rubrene). 
A finite $\xi$ causes an energy unbalance between the initial and
final hopping sites (the double-well potential
$E_{ad}(Q)$  becomes asymmetric) so that the hopping rate changes to
$\rm{w}(\xi)=\rm{w}_0 e^{ -\xi/2k_BT}$. 
Now each carrier diffuses with a different hopping rate, determined by its own
environment. Correspondingly, the mobility of the sample is obtained as the
statistical average $\langle \rm{w}(\xi)\rangle$ over all the possible
values of the local electronic potential $\xi$.

Since the polaronic barrier sets the dominant energy scale, due to the
high polarizability of the gate dielectric Ta$_2$O$_5$ used in our devices, 
we always have  $\xi \ll \Delta$ in the explored density range. 
As a result, correlations between
subsequent hops can be neglected to a first approximation, 
which  corresponds to
replacing $\langle \rm{w}(\xi) \rangle \to \rm{w}(\langle\xi
\rangle)$. This yields the leading term of the
mobility in the presence of electron-electron interactions as 
\begin{equation}
  \label{eq:intmob}
  \mu(T)=\mu_P(T) e^{ -\langle\xi \rangle/2k_BT}.
\end{equation}
The density dependent quantity $\langle\xi \rangle$ represents the
average extra energy
cost for hopping from site to site, induced by the
long-range Coulomb interactions between the carriers.
As the comparison between theory and experiments will show (see below), it is this dependence of $\mu$ on the density (through $\langle\xi \rangle$) that 
causes the saturation  of $I_{sd}$ observed in Fig. 1.

In the linear response regime the calculation of $\langle \xi \rangle$
follows, through Eq. (\ref{eq:xi}), from the
statistical distribution of the occupation numbers
$\{n_j\}$ {\it constrained} to $n_1=1$ (site 1 is initially occupied by the
hole, which then hops to site 2). Such constraint reflects the
fact that the relaxation
of the remaining carriers occurs on a time scale $\sim \rm{w}_0^{-1}$, much
longer than that of the hopping event under consideration, $\sim
\omega_s^{-1}$, and can be neglected.  In this case, the occupation
numbers $\{n_j\}$ are given, by definition, by  the
pair distribution function $g^{(2)}(R_{1j})$ of the electronic
system at equilibrium \cite{Hansen}.  Furthermore, due to the diffusive
nature of their hopping motion (transport is thermally activated),
the small polarons behave  as
classical interacting particles. The two-body correlations of such a classical
fluid of charged particles are determined solely by the Coulomb interactions,
and are therefore equivalent to those of a classical two-dimensional plasma.
Observing that in the explored density range 
the mean inter-particle distance is always larger than
the lattice periodicity,
we can  use for $g^{(2)}$ the known pair distribution
function of the 2D plasma in the continuous limit. 
Correspondingly, we can replace the discrete sum
in Eq. (\ref{eq:xi})  by an integral, which finally yields
\begin{equation}
  \label{eq:xiavg}
  \langle\xi \rangle= n \int d {\bf r} \; [\bar{V}({\bf r}+
  {\bf R}_{12})-\bar{V}({\bf r})] \; g^{(2)}(r)\equiv \frac{\pi}{2} na^2 k_B T
  F(\Gamma). 
\end{equation}
In the above equation, the universal 
function $F(\Gamma)$ is an intrinsic property of the two-dimensional
electron plasma. It depends on a single 
dimensionless parameter $\Gamma$ that measures the ratio
of the Coulomb interactions to the thermal energy. $F(\Gamma)$  
can be evaluated directly by  using  the data of $g^{(2)}(r)$ available
from extensive Monte Carlo simulations  \cite{MC}.  
In the range $1<\Gamma<20$, we find
$F(\Gamma)=1+0.85\Gamma$ to within $1\%$ accuracy. 
In the present FET geometry, the plasma interaction parameter  is
given by $\Gamma=2\sqrt{\pi n}e^2/(k_BT)/(\epsilon_s+\kappa)$, and
can attain the value $\sim 9$ at the lowest temperatures/highest 
$V_g$, placing our devices in the range of weak to moderate interactions
(electron crystallization is expected at $\Gamma=125$).

\section{Discussion}

We are now in a position to compare the experimental data with our
calculations. 
By inserting Eq.(\ref{eq:xiavg}) into Eq.(\ref{eq:intmob}), and using the
definition of $\Gamma$, we obtain the explicit functional dependence
of the mobility $\mu$ on density $n=C(V_g-V_{th})/e$  and
temperature. The $I_{sd}-V_g$ curves can then be calculated from
$I_{sd}= A |V_g-V_{th}| e^{-\langle\xi \rangle/2k_BT}$,
with $A= \mu_P C V_{sd} W/L$ having the dimensions of conductance  
(here $W$ and $L$ are the width and the length of the channel, $V_{sd}$ the
applied source-drain voltage, and $e$ the electron charge). To analyze the 
data, we fix the temperature and fit the $I_{sd}-V_g$ curve using 
$\mu_P$ and $V_{th}$ as fitting parameters. The polaron
microscopic parameters are obtained by analyzing the
temperature dependence  of $\mu_P$ extracted in this way.

\begin{figure}
  \centering
  \includegraphics[width=80mm]{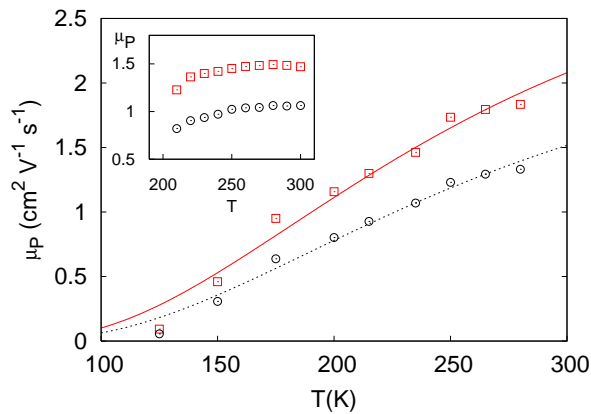}
   \caption{Polaron mobility extracted from the data of Fig. 1a. Red
     squares are obtained by fitting the data with the theory for
     interacting polarons described in Sec. 3, while black circles
     correspond to linear fits of the $I_{sd}-V_g$
     curves restricted to the low density regime, as done in
     Ref.\cite{NatMat}.  The lines are best fits to the polaronic thermally
     activated behavior, yielding respectively $\Delta=53 meV$ and
     $\Delta=55 meV$ (see text). The inset is the polaron mobility for the
     sample of Fig. 1b (units are the same as in the main panel; the
     temperature range is too small to extract reliable values for $\Delta$).} 
  \label{fig:parameters}
\end{figure}

The values of $\mu_P$ extracted for the different devices from the
theory of interacting polarons are reported in
Fig. \ref{fig:parameters} (red squares). They are 
close to the ones that one would obtain from linear fits of the $I_{sd}-V_g$
curves in the low density regime (the conventional definition of the
mobility; data shown as black circles in Fig. \ref{fig:parameters}) 
and they exhibit the same trend.  Indeed, by comparing the temperature 
dependence of the fitted $\mu_P$  to the theoretical relation 
$\mu_P(T)=(ea^2\omega_s /2\pi k_B T)\exp (-\Delta/k_BT)$ we
obtain, for the device of Fig. 1a, 
the following values of the polaron parameters: 
$\omega_s/(2\pi)=390 cm^{-1}$ and $\Delta=53 meV$. 
These values are comparable to those obtained by assuming a 
linear $I_{sd}$ vs. $V_g$ dependence [$\omega_s/(2\pi)=315 cm^{-1}$ and
$\Delta=55 meV$].  This shows that the results of the interpretation
based on the theory with interactions is compatible with the analysis
performed by only looking at the low density part of the $I_{sd}-V_g$
curves, and that the effect of the interactions is determined by the
carrier density only (i.e., without the need to include any additional
parameter).    
For the second device (inset of Fig. \ref{fig:parameters}), 
a precise quantitative analysis of the temperature
dependence of the  mobility is prevented by the restricted  temperature range 
of the available data, and it is not possible to extract reliable values 
for the parameters $\omega_s/(2\pi)$ and  $\Delta$. Still, the fitted
values of $\mu_P$ and $V_{th}$ are well defined for all the curves of 
Fig. 1b, thus allowing  for an accurate analysis of the density 
dependence of the mobility at each given temperature.

The values of $V_{th}$ obtained from the
interacting theory also follow the same trend as
those that one obtains by extrapolating the linear part of the
$I_{sd}-V_g$ curves to zero current, i.e. the usual definition of
$V_{th}$. The two estimates differ by a systematic offset of 
$\sim 7$ V, which is approximately the same 
for all samples and at all  temperatures. 
Such a deviation does not have a particular physical significance, 
because the exact absolute value of $V_{th}$ defined in the usual
way does not have a precise physical meaning.

The continuous lines in Fig. 1.a and b are the results of the
theoretical fits of the $I_{sd}-V_g$ curves, using the theory for
interacting polarons.   
The theory reproduces the saturation of the
source-drain current at high carrier density and the quantitative
agreement with the data is remarkable, at all temperatures and in both
the devices analyzed. 
From Eq.(\ref{eq:intmob}) we see that theory predicts an 
activation energy which increases with increasing density (via the
term $\langle\xi \rangle/2$). This is illustrated in the 
 inset of Fig. 1a, where the circles are the experimental values
 extracted from the data and the continuous line is the theoretical 
curve computed using Eq.(\ref{eq:xiavg}), again
without any new free parameters. Also here the 
agreement is very satisfactory, and at small $n$ we recover the
value given by the analysis of the linear regime in terms of 
non-interacting polarons \cite{NatMat}.  

\begin{figure}
  \centering
  \includegraphics[width=80mm]{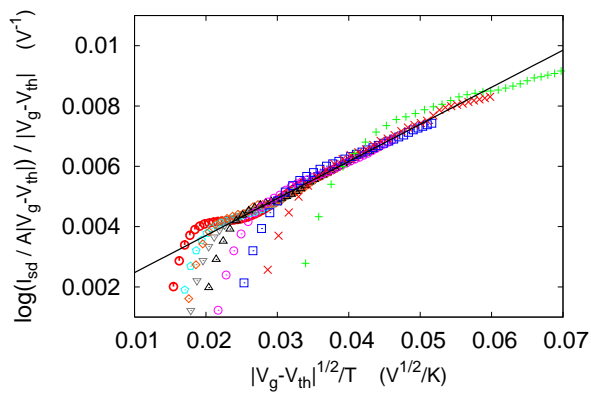}
   \caption{Scaling plot of the data of Fig. 1.a.    (the same symbols
     have been used). The full line corresponds to the prediction of the 2D 
classical plasma.} 
  \label{fig:scaling}
\end{figure}

Notably, the theoretical derivation presented above implies that
the density induced increase of the hopping activation barrier [cf. Eq.
 (\ref{eq:intmob})] is a scaling function of  the
parameter $\Gamma$ of the interacting plasma. This scaling behavior is checked
in Fig. \ref{fig:scaling}, where we use the experimental data to directly plot the quantity $\log(I_{sd}/A|V_g-V_{th}|)/|V_g-V_{th}|$,  which is proportional to the universal function $F(\Gamma)$ (see Eqs. (\ref{eq:intmob}) and
(\ref{eq:xiavg})),  versus $\sqrt{|V_g-V_{th}|}/T\propto \Gamma$.
In terms of these variables and at sufficiently high gate voltages, the data
do indeed tend to collapse on a single linear curve corresponding
to  $F(\Gamma) = 1 + 0.85 \Gamma $ 
(full line in the figure).
Considering the high sensitivity to details of such scaling plot, 
the agreement between theory and experiment is satisfactory
It confirms the validity of the assumptions underlying our derivation,
giving a strong indication
 that the anomalous behavior of the electrical  characteristics 
observed at  finite carrier concentrations is indeed due to the
long-range Coulomb interactions between the carriers.

\section{Conclusion}
We have studied electronic transport through rubrene
single-crystal FETs in the  high-density regime and found new
unexpected phenomena, such as a saturation of the current 
$I_{sd}$ vs. $V_g$ and
an  increase of the activation energy with increasing density. 
The experimental data are accurately
reproduced by a theoretical analysis based on interacting small Fr\"ohlich
polarons, which naturally extends previous studies in the low-density
regime to include the effect of the long-range Coulomb repulsion between
charge carriers.  Remarkably, the quantitative description of the
high-density regime does not require the introduction of any new
parameter, as the effect of the interactions is fully determined by the
(known) density of charge carriers. 
Our results demonstrate that, by devoting sufficient effort to control
the experimental systems, single-crystal organic field-effect
transistors do allow a detailed quantitative study of the intrinsic
transport properties of organic FETs.

\ack
We thank R.W.I. de Boer and N. Iosad for experimental help and 
NWO and Nanoned for financial support.

\appendix

\section{Calculation of the polaron radius}

The theoretical analysis developed in the text treats the charge
carriers as small Fr\"ohlich polarons, having dimensions comparable to
the lattice spacing. To demonstrate that this is the case,
in this Appendix we evaluate the radius of the polarons that form at a
rubrene/Ta$_2$O$_5$ interface.
To this aim we solve the adiabatic
model Eq. (\ref{eq:H}) variationally for a single hole in the HOMO band, and
evaluate the polaron radius $R_P$ as a function of the electron-phonon
coupling strength. The calculations indeed show that $R_P$ is of the order of
the lattice spacing. 

The electron-phonon interaction $g_{ij}$ is obtained from the Fourier
transform of the electron-phonon matrix element $M_q=M_0
e^{-qz}/\sqrt{q}$ \cite{MoriAndo,NatMat}. 
Here  $M_0^2=2\pi \hbar \omega_s e^2 \beta/S$, with $S$
the total surface of the device, and $z$ the distance of
the conducting layer to the polar dielectric. 
The parameter $\beta=(\epsilon_s-\epsilon_\infty)/(\epsilon_s+\kappa)/
(\epsilon_\infty+\kappa)$ is a combination of the known dielectric
constants of the gate dielectric and of the organic semiconductor,
which determines the strength of the electron-phonon coupling.
In principle, the interaction $g_{ij}$ defined above
must be cut off at short distances to account for the discreteness of the
rubrene lattice. A precise prescription 
to carry out this procedure will be reported elsewhere. For the
present purposes  it suffices to say that
the overall behavior of the polaron radius does not depend  
on the particular choice of the short-distance cutoff.

As customary for lattice models, we introduce 
the dimensionless electron-phonon coupling
\begin{equation}
  \label{eq:lambda}
  \lambda=\frac{E_P}{D},
\end{equation}
defined as the energy of a fully localized polaron in units of the half
bandwidth $D\propto t$. The energy $E_P$ can be written in general as:
\begin{equation}
  \label{eq:Ep}
  E_P=\frac{\sum_{R_i}g_{0i}^2}{\hbar \omega_s},
\end{equation}
wher the sum spans all the lattice sites $R_i$. 
The model is solved by taking the following gaussian
trial wavefunction: 
\begin{equation}
  \label{eq:trial}
  \phi(R_i)= \frac{1}{\mathcal{N}^{1/2}} \exp (-\alpha^2 R_i^2/2)
\end{equation}
with $\mathcal{N}$  a normalization factor.
The parameter $\alpha$ is obtained after minimization of
the ground state energy. The polaron radius is then
given by
\begin{equation}
  \label{eq:radius}
  R_P=\left[ \sum_{R_i} \phi^2(R_i) R_i^2\right]^{1/2}
\end{equation}
which tends to $1/\alpha$ in the large polaron limit (small $\alpha$,
$R_P\gg a$).
The results for $R_P$ at different values of the
electron-phonon coupling are summarized in Table \ref{tab:radius}.
As expected, the polaron radius becomes comparable to the lattice
spacing around $\lambda\sim 1$, i.e. when the polaron energy is
of the order of the half bandwidth. 

\begin{table}[htbp]
  \centering
  \begin{tabular}{|c|c c c c c c c c c|}
 \hline \hline
$\lambda$   & 0.1 & 0.2 & 0.4 & 0.7 & 1.0 & 1.2 & 1.4 & 1.6 &2\\
\hline
  $R_P/a$ & 6.0& 3.7 & 2.3 & 1.6 & 1.2 & 0.9 & 0.5 & 0.4 & 0.3\\
\hline \hline
  \end{tabular}
  \caption{Polaron radius in units of the lattice spacing $a$ 
for different values of the dimensionless electron-phonon coupling $\lambda$.}
  \label{tab:radius}
\end{table}

Performing the sum in Eq. (\ref{eq:Ep}) with the electron-phonon
coupling parameters  appropriate to our devices ($\beta=0.1$ and $z=2\AA$), 
we obtain $E_P=170 meV$.
The width of the HOMO band in the rubrene crystal has been evaluated through a
semi-empirical {\it ab initio} calculation in Ref. \cite{DaSilva} as
$2D=341 meV$. Such  value should be considered as an upper bound to the
actual bandwidth, which was shown in Ref. \cite{Bussac} to be 
sizably reduced by the effects of molecular polarization. A further
effective reduction of the bandwidth can arise from thermal
fluctuations, as considered in Ref. \cite{Troisi}.
From the above arguments and from 
the definition Eq. (\ref{eq:lambda}) we conclude that  
$\lambda > 1$ which, as can be seen from Table
\ref{tab:radius}, corresponds to a small polaron.



\section*{References}
\begin{thebibliography}{99}

\bibitem{RevModPhys} M.E. Gershenson, V. Podzorov, and A.F. Morpurgo, Rev. Mod. Phys. {\bf 78}, 973 (2006)
\bibitem{Sundar} V.C. Sundar {\it et al.}, Science {\bf 303}, 1644 (2004)
\bibitem{Podzorov1} V. Podzorov {\it et al.}, Phys. Rev. Lett. {\bf 93}, 086602 (2004)
\bibitem{Podzorov2} V. Podzorov {\it et al.}, Phys. Rev. Lett. {\bf 95}, 226601 (2005)
\bibitem{Basov} M. Fischer, et al, Appl. Phys. Letters 89, 182103
  (2006) \\ Z.Q. Li {\it et al.}, Phys. Rev. Lett. {\bf 99}, 016403 (2007)
\bibitem{NatMat} I. N. Hulea et al., Nature Materials 5, 982 (2006)
\bibitem{Kirova}N. Kirova, M. N. Bussac, Phys. Rev. B 68, 235312 (2003)
\bibitem{deBoer} R.W.I. de Boer, T.M. Klapwijk, and A.F. Morpurgo, App. Phys. Lett. {\bf 83}, 4345 (2003)
\bibitem{Review} R.W.I. de Boer {\it et al.}, Phys. Stat. Sol. A {\bf 201}, 1302 (2004)
\bibitem{leakage} R.W.I. de Boer {\it et al.}, App. Phys. Lett. {\bf 86}, 032103 (2005)
\bibitem{Shimotani} H. Shimotani et al, Appl. Phys. Letters 89, 203501
  (2006) 
\bibitem{Frisbie} M.J. Panzer, C.D. Frisbie, J. Am. Chem. Soc. {\bf 129}, 6599 (2007)
\bibitem{Troisi} A. Troisi, G. Orlandi, Phys. Rev. Lett. {\bf 96}, 
086601 (2006)
\bibitem{polcryst} S. Fratini, P. Qu\'emerais, Eur. Phys. J. {\bf B
    14}, 99 (2000)




\bibitem{MoriAndo} J. Sak, Phys. Rev. B 6, 3981 (1972);
N. Mori, T. Ando, {\it ibid} 40, 6175 (1989)

\bibitem{Verbist}     G. Verbist, M. A. Smondyrev, F. M. Peeters, and
  J. T. Devreese, Phys. Rev. B 45, 5262 (1992) 

\bibitem{LangFirsov}I. G. Lang, Yu. A. Firsov, Sov. Phys. Solid State
  9, 2701 (1968) 

\bibitem{Chandra} S. Chandrasekhar, Rev. Mod. Phys. {\bf 15}, 1 (1943)


\bibitem{Hansen} J. P. Hansen and I. R. McDonald, {\it Theory of
    Simple Liquids}  (Academic, London, 1976).

\bibitem{MC} H. Totsuji, Phys. Rev {\bf A 17}, 399 (1978)


\bibitem{DaSilva} D. A. da Silva Filho,  E.-G. Kim, J.-L. Br\'edas,  Adv. Materials 17, 1072 (2005)

\bibitem{Bussac} M. N. Bussac, J. D. Picon, L. Zuppiroli,
  Europhys. Lett. 66, 392 (2004)

\end{thebibliography}
\end{document}